# Basic Theoretical Model and Sampling Criteria for Time-Interleaved Photonic Analog-to-Digital Converters

Feiran Su, Guiling Wu, Weiwen Zou, *Senior Member*, *IEEE*, and Jianping Chen

*Abstract*—In this paper, we present a basic model for time-interleaved photonic analog-to-digital converter (TIPADC) and analyze its linear, nonlinear, and noise performance. The basic operation mechanism of TIPADC is illustrated and the linear performance is analyzed in frequency domain. The mathematical expressions and the output of the system are presented in each processing step. We reveal that photonic sampling folds the whole spectrum of input signal in a narrow band, which enables the analog bandwidth of a TIPADC to be much higher than the bandwidth of back-end electronics. The feasible regions of the system is obtained in terms of system frequency response, and a set of sampling criteria determining the basic requirements to the system are summarized for typical applications. The results show that the global minimum feasible bandwidth of back-end electronics is the half of single channel sampling rate when the bandwidth of photonic sampling pulse is comparable to input signal frequency. The amplitude and phase noise of photonic sampling pulse train, the thermal noise and shot noise of photodetection, and the quantization noise and aperture jitter noise of EADCs are investigated. The nonlinearity of TIPADC is also analyzed. The effect of different factors, such as the single channel sampling rate, the average optical power and the bandwidth of back-end electronics, on the effective number of bits of TIPADC is qualitatively and quantitatively discussed. The results explicitly show that TIPADC can perform high fidelity conversion to ultra-high speed analog signal and overcome the major performance limitations existing in traditional ADCs such as aperture jitter and bandwidth. The criteria deduced will be an effective supplement to the generalized sampling theorem in designing TIPADC.

## NOMENCLATURE

| | |
|---|---|
| $v_I(t)$ | Input radio frequency (RF) signal. |
| $N$ | Number of channels in a TIPADC. |
| $T_S$ | Single channel sampling period. |
| $f_S$ | Single channel sampling rate, $f_S = 1/T_S$. |
| $\Omega_S$ | Single channel sampling angular frequency. |
| $P_A$ | Average power of the photonic sampling pulse train in a channel. |
| $p_n(t)$ | Photonic sampling pulse train in the $n$th channel. |
| $p_S(t)$ | Normalized photonic sampling pulse shape. |
| $\Delta p_n(t)$ | Noise of the photonic sampling pulse train. |
| $\Delta v_n(t)$ | Noise referred to the input of electronic quantization. |
| $t_M(v)$ | Transmittance of electro-optical modulator (EOM). |
| $h_M(t)$ | Small-signal impulse response of EOM. |
| $R_D$ | Responsibility of photodetector. |
| $R_L$ | Transimpedence of photodetector. |
| $h_E(t)$ | Impulse response of back-end electronics. |
| $f_n[k]$ | Digital filter in the $n$th channel. |
| $v_O[k]$ | Output digital samples. |
| $V_I(\Omega)$ | Spectrum of $v_I(t)$. |
| $P_n(\Omega)$ | Spectrum of $p_n(t)$. |
| $P_S(\Omega)$ | Spectrum of $p_S(t)$. |
| $H_E(\Omega)$ | Frequency response of back-end electronics. |
| $H_A(\Omega)$ | Equivalent frequency response of system. |
| $H_P(\Omega)$ | Equivalent frequency response of system to the phase noise of the photonic sampling pulse train. |
| $H_J(\Omega)$ | Equivalent frequency response of system to the aperture jitter noise of EADC. |
| $G_E$ | Gain in the passband of $H_E(\Omega)$. |
| $\beta_P$ | Bandwidth of $P_S(\Omega)$ normalized by $f_S$. |
| $\beta_E$ | Bandwidth of $H_E(\Omega)$ normalized by $f_S$. |
| $\chi_E$ | Center frequency of $H_E(\Omega)$ normalized by $f_S$. |

Manuscript received February 13, 2015. This work was supported in part by the National Natural Science Foundation of China (61127016), the 973 Program (2011CB301700), SRFDP of MOE (Grant No. 20130073130005). (Corresponding author: Guiling Wu).

The authors are with State Key Laboratory of Advanced Optical Communication Systems and Networks, Department of Electronic Engineering, Shanghai Jiao Tong University, Shanghai 200240, China (e-mail: feiran.su@gmail.com; wuguiling@sjtu.edu.cn; wzou@sjtu.edu.cn; jpchen62@sjtu.edu.cn).

| | |
|---|---|
| $\chi_U$ | Upper cutoff frequency of system normalized by $f_S$. |
| $\chi_L$ | Lower cutoff frequency of system normalized by $f_S$. |
| $\eta$ | Order of Butterworth filter. |

## I. INTRODUCTION

Photonic analog-to-digital converter (PADC) has been proposed as a promising alternative to the electronic ADC (EADC). The unique advantages of photonics, such as ultra-low timing jitter, ultra-high speed, and wide bandwidth processing, can overcome the bandwidth and resolution limitation [1]-[3]. In the schemes proposed so far, the time-interleaved PADC (TIPADC) is one of the most practically feasible methods [4]-[5]. It employs time-interleaved optical sampling and back-end electronic quantization. The reason employing electronic quantization lies in the fact that there has been no photonic quantization technique with high effective number of bits (ENOB) [6]-[8] for most of practical applications. Though several experimental systems of TIPADC have been demonstrated [9]-[12], they focus mainly on the photonic issues. A basic theoretical model, which can describe the systematic operation mechanism of TIPADC including back-end electronics and provide guidance for its optimal design, is still an open issue.

In this paper, we present a basic theoretical model of TIPADC and analyze its linear, nonlinear, and noise performance. Based on the model, the signal sampling and reconstruction are depicted in detail. The linear performance is analyzed and the system frequency response is derived. The feasible regions of system parameters are disclosed and a set of sampling criteria for typical applications are summarized in terms of system frequency response. Based on the presented model and its linear analysis, we investigate the impact of the inter-symbol interference (ISI) in the photodetection and electronic quantization. Numerical simulations are carried out on a 32-channel TIPADC with 1Gs/s sampling rate per channel and the results agree well with the theoretical analysis. Furthermore, we derive the output powers of major noises and nonlinear distortion in TIPADCs, including the amplitude and phase noise of photonic sampling pulse train, the thermal and shot noise of photodetection, the quantization and aperture jitter noise of EADCs, and the modulator nonlinearity. The effects of system parameters and noises on ENOB are discussed.

The rest of this paper is organized as follows. Section II describes the basic model of TIPADC. Section III presents its linear performance. Section IV demonstrates its noise and nonlinear performance. Conclusions are remarked in Section V.

## II. THE BASIC MODEL OF TIPADC

A typical *N*-channel TIPADC is schematically shown in Fig. 1 (a). Its operation can be divided into four steps: i) photonic sampling and demultiplexing, ii) photodetection, iii) electronic quantization, and iv) digital processing. The functionality of each step is as follows. The photonic sampling pulse train from an optical pulse generator is modulated by a radio frequency (RF) signal through an electro-optical modulator (EOM) such as a Mach-Zehnder modulator (MZM), and demultiplexed to *N* parallel lower speed channels via wavelength division multiplexing (WDM) and/or optical time-division multiplexing (OTDM). The photodetection converts the optical signals in all parallel channels into electric signals. The EADCs quantize them into digital samples. The digital processing combines them into the digital form of the input RF signal.

According to the above operation mechanism, a TIPADC can be generally modeled as Fig. 1 (b). The definitions of the symbols in the figure are listed in Nomenclature. In order to simplify the analysis, we consider all the sampling channels are matched, i.e. different channels have the same gain and uniform delay. In fact, the mismatches can be calibrated to a certain extent via algorithms [13]. Hence such assumption does not affect the generality of the model.

### A. Photonic Sampling and Demultiplexing

We assume for analysis simplicity that the nonlinear and polarization effects in optical fiber and devices are negligible. In fact, the average optical power of each photonic sampling pulse train is low enough [10]-[11] and polarization effect can be alleviated via polarization controllers. Therefore the above assumption is valid. In this case, each channel in the multichannel photonic sampling model is independent.

In a TIPADC, there exist relative delays between the photonic sampling pulse trains in different channels. In the proposed model, we consider these delays as the advances of input RF signal to corresponding channels. Thus, the photonic sampling pulse train in the *n*th channel can be expressed as [14]-[15]:

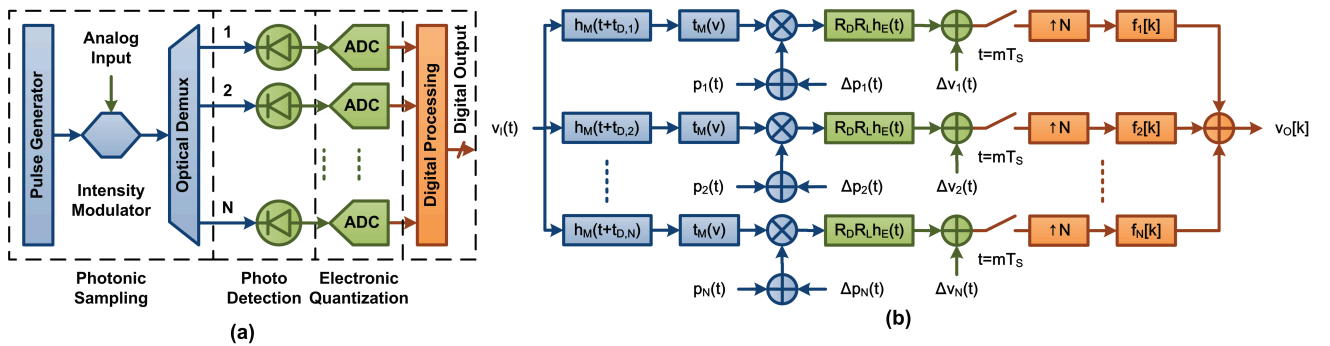

Fig. 1. Schematic structure (a) and the basic theoretical model (b) of TIPADC.

$$p_n(t) = P_A \sum_{m=-\infty}^{\infty} p_S(t - mT_S). \tag{1}$$

The effective signal, $v_{M,n}(t)$, that modulates $p_n(t)$ is determined by the small-signal impulse response of EOM and given by:

$$v_{M,n}(t) = v_I(t) * h_M(t + t_{D,n}), \tag{2}$$

where $t_{D,n} = (n-1)T_S/N$ is the time advance introduced by the delay relative to the first channel.

After modulation, the sampled pulse train can be expressed as [14]-[15]

$$p_{M,n}(t) = t_M(v_{M,n}) p_n(t) \tag{3}$$

Typically, $t_M(v)$ is a nonlinear function of $v_{M,n}(t)$ and causes nonlinear distortion. For Mach-Zehnder intensity modulator, we have

$$t_M(v_{M,n}) = \frac{1}{2} + \frac{1}{2} \cos\left[\frac{\pi}{V_\pi}(v_{M,n} + v_B)\right]. \tag{4}$$

where $V_\pi$ is the half wave voltage. For maximum modulation efficiency, MZM should be biased at $v_B = V_\pi/2$.

The amplitude and phase noises related to the photonic sampling pulse train are modeled as additive noise sources expressed by $\Delta p_n(t)$ in the model. Then the noise after modulation can be expressed as

$$\Delta p_{M,n}(t) = t_M(v_{M,n}) \Delta p_n(t). \tag{5}$$

*B. Photodetection and Electronic Quantizationp*

In the circumstance that the optical signal power on each device (such as photodetector and electrical amplifier) does not exceed the threshold, as is the usual case in a practical TIPADC, the nonlinearity of devices can be ignored. The frequency responses of devices in photodetection and electronic quantization can be accounted into a single function $h_E(t)$ since they are linear and cascaded. Thus, the photodetected signal can be expressed as

$$v_{D,n}(t) = R_D R_L h_E(t) * p_{M,n}(t), \tag{6}$$

where $R_D$ and $R_L$ is the responsibility and transimpedance gain of photodetector, respectively.

The noises in these steps include the intrinsic thermal noise and shot noise in photodetection, and the quantization noise and aperture jitter noise in the electronic quantization. All of them are modeled as additive noises and expressed by $\Delta v_n(t)$ together. Note that the quantization noise and aperture jitter noise have been referred to the input of the electronic quantization. Hence, the total noise can be expressed as

$$\Delta v_{D,n}(t) = R_D R_L h_E(t) * \Delta p_{M,n}(t) + \Delta v_n(t). \tag{7}$$

EADCs sample the detected signals in all channels simultaneously. The samples and noise after quantization can be written as

$$v_{Q,n}[k] = v_{D,n}(kT_S) \tag{8}$$

$$\Delta v_{Q,n}[k] = \Delta v_{D,n}(kT_S). \tag{9}$$

*C. Digital Processing and the Output of System*

Digital processing can be modeled as the combination of up-sampler, digital filter $f_n[k]$ and adder. The up-sampler increases the sampling rate with a factor of $N$ by adding $N-1$ zeros between two consecutive samples of $v_{Q,n}[k]$:

$$v_{U,n}[k] = \begin{cases} v_{Q,n}[k/N], & k = 0, \pm 1, \cdots \\ 0, & others \end{cases}. \tag{10}$$

The digital filter $f_n[k]$ is used to represent the linear processing such as delay or matching among channels. In our model we just consider it as an ideal delayer:

$$f_n[k] = \delta[k - n + 1]. \tag{11}$$

Then the adder outputs the combined sequence $v_O[k]$:

$$v_O[k] = \sum_{n=1}^{N} f_n[k] * v_{U,n}[k]. \tag{12}$$

There is no additional noise in this step since it is a digital processing.

### III. THE LINEAR PERFORMANCE OF TIPADC

In addition to the noise and nonlinearity, the sampling result of TIPADC is affected by the system parameters that bring about linear distortion. In this section, we characterize the linear performance of TIPADC by deriving the equivalent frequency response from the basic model.

*A. The Linear Component of Output and the Equivalent Frequency Response of System*

TIPADC can be linearized and analyzed in the frequency domain. The spectrum of the photonic sampling pulse train in the *n*th channel can be derived directly from (1):

$$P_n(\Omega) = 2\pi P_A \sum_{m=-\infty}^{\infty} c_m \delta(\Omega - m\Omega_S), \tag{13}$$

where the coefficients $c_m$ are determined by $P_S(m\Omega_S)$ with $c_0 = 1$ due to the normalization of $p_S(t)$.

To extract the linear component of $p_{M,n}(t)$, we assume the amplitude of input RF signal is sufficient small, i.e., $v_{M,n}(t) \ll 2V_\pi$. The small signal approximation of $p_{M,n}(t)$ can be derived from (3) and (4) for quadrature-biased MZMs:

$$p_{M,n}(t) \approx [a_0 + a_1 v_{M,n}(t)] p_n(t), \tag{14}$$

where $a_0 = 0.5$, $a_1 = -\pi/2V_\pi$. Then the spectrum of the modulated photonic pulse train can be expressed as

$$P_{M,n}(\Omega) = a_0 P_n(\Omega) + a_1 P_A \sum_{m=-\infty}^{\infty} c_m V_{M,n}(\Omega - m\Omega_S) \tag{15}$$

where $V_{M,n}(\Omega)$ is the spectrum of $v_{M,n}(t)$ and is related to EOM's small-signal frequency response, $H_M(\Omega)$, via:

$$V_{M,n}(\Omega) = V_I(\Omega) H_M(\Omega) \exp(j\Omega t_{D,n}). \quad (16)$$

After photodetection, the spectrum of the voltage signal to be digitized can be derived as

$$V_{D,n}(\Omega) = a_0 R_D R_L H_E(\Omega) P_n(\Omega) + V_{D1,n}(\Omega), \quad (17)$$

where

$$V_{D1,n}(\Omega) = a_1 P_A R_D R_L H_E(\Omega) \sum_{m=-\infty}^{\infty} c_m V_{M,n}(\Omega - m\Omega_S). \quad (18)$$

After digitization, the first term of (17) is aliased to a DC term. Hence it can be eliminated digitally, and the spectrum of the remaining component can be derived as follows, according to Poisson's summation formula:

$$V_{Q,n}(\omega) = \frac{1}{T_S} \sum_{l=-\infty}^{\infty} V_{D1,n}\left(\frac{\omega}{T_S} - \frac{2\pi l}{T_S}\right) = \frac{a_1 P_A R_D R_L}{T_S}$$
$$\cdot \sum_{l=-\infty}^{\infty} \sum_{m=-\infty}^{\infty} c_m H_E\left(\frac{\omega}{T_S} - \frac{2\pi l}{T_S}\right) V_{M,n}\left[\frac{\omega}{T_S} - \frac{2\pi}{T_S}(l+m)\right] \quad (19)$$

The input and output of up-sampling are related as

$$V_{U,n}(\omega) = V_{Q,n}(\omega N). \quad (20)$$

Then the spectrum of system output can be derived as

$$V_O(\omega) = \sum_{n=1}^{\infty} V_{U,n}(\omega) \exp[-jw(n-1)]$$
$$= \frac{N}{T_S} \sum_{l=-\infty}^{\infty} H_A\left(\frac{\omega N}{T_S} - \frac{2\pi l N}{T_S}\right) V_I\left[\frac{\omega N}{T_S} - \frac{2\pi l N}{T_S}\right], \quad (21)$$

where $H_A(\Omega)$ is the equivalent frequency response of system and is given by

$$H_A(\Omega) = a_1 P_A R_D R_L H_M(\Omega) \sum_{m=-\infty}^{\infty} c_m H_E(\Omega + m\Omega_S), \quad (22)$$

Correspondingly, the system impulse response is

$$h_A(t) = a_1 R_D R_L p_n(-t) h_E(t) * h_M(t). \quad (23)$$

From (21), we can see that the system is equivalent to a single ADC sampling $v_I(t)$ uniformly at a rate of $Nf_S$ with the equivalent frequency response of $H_A(\Omega)$. This implies that the TIPADC shown in Fig. 1 can be referred to an *N*-channel generalized sampling system [16]. (22) indicates that the analog bandwidth of TIPADC can be much higher than the bandwidth of back-end electronics because of the weighted summing introduced by the photonic sampling step. The photonic sampling pulse train has more functionality than the traditional sampling clocks and photonic sampling realized via EOM in principle involves mixing manipulation. As will be demonstrated in the following section, such manipulation folds the whole spectrum of the input signal in a narrow band, resulting in the weighted summing. In contrast, the system analog bandwidth in a traditional time-interleaved ADC, where there is no front-end sampling step, is limited by the bandwidth of back-end electronics [16]. Moreover, the frequency response of TIPADC is not related to the number of channels, *N*, as shown in (22). That means increasing *N* can only increase the sampling rate of system but cannot extend the system bandwidth. In order to detect a wideband signal, both system analog bandwidth and sampling rate should be sufficient. When *N* is 1, TIPADC degenerates to subsampling photonic ADC [17]. In this case, there is no effect of channel mismatches on ENOB. The bandwidth of input RF signals, however, is limited in order to avoid aliasing since the sampling rate of the system decreases to $f_S$.

We take a 4-channel TIPADC as an example to illuminate the signal spectrum in each processing step. In the photonic sampling step, $P_n(\Omega)$ is modulated with an input RF signal $V_I(\Omega)$ whose bandwidth is within half of the system sampling rate, i.e., $0.5 \times 4f_S$. It can be seen that photonic sampling folds all components of $V_I(\Omega)$ into the frequency bands with bandwidth of $f_S$, where aliasing occurs when the bandwidth of $V_I(\Omega)$ is larger than $0.5 f_S$, as shown in Fig. 2 (a). For a bandwidth limited lowpass $H_E(\Omega)$, the photodetection receives the signal in the passband of $H_E(\Omega)$, i.e. $V_{D,n}(\Omega)$, which contains all components of the input signal, as shown in the top sub-figure of Fig. 2 (b). In the electronic quantization step it is converted to digital form $V_{Q,n}(\omega)$ and is aliased once again (see the middle sub-figure of Fig. 2 (b)). In the digital processing step, the output sequence $V_O(\omega)$ is obtained by processing $V_{Q,n}(\omega)$ in each channel. The final combination results in the cancellation of aliases and $V_O(\omega)$ is only linearly distorted by the system response in comparison with $V_I(\Omega)$, as shown in the bottom of Fig. 2 (b). Fig. 2 (c) shows the case where $H_E(\Omega)$ is bandpass. The result is similar to those shown in Fig. 2 (b).

*B. The Criteria of Reconstructing a Wideband Signal in Terms of Frequency Response*

From (21), we can see that only the input RF signal lying in the passband of $H_A(\Omega)$ can be reconstructed. For a wideband TIPADC, it implies that the bandwidth of $H_M(\Omega)$ should be sufficient. To analyze the criteria determined by other factors, we assume $H_M(\Omega) = 1$ in this section.

As (22) shows, $H_A(\Omega)$ is the weighted sum of replicas of $H_E(\Omega)$. The *m*th replica of $H_E(\Omega)$ is shifted to $-m\Omega_S$ with weight of $c_m$. The passband of $H_A(\Omega)$ is the frequency range where at least one of the weighted replicas of $H_E(\Omega)$ has non-zero value and its bandwidth is related to the overlapping of replicas. Therefore the passband can be derived by analyzing the existence and the overlapping of the replicas of $H_E(\Omega)$.

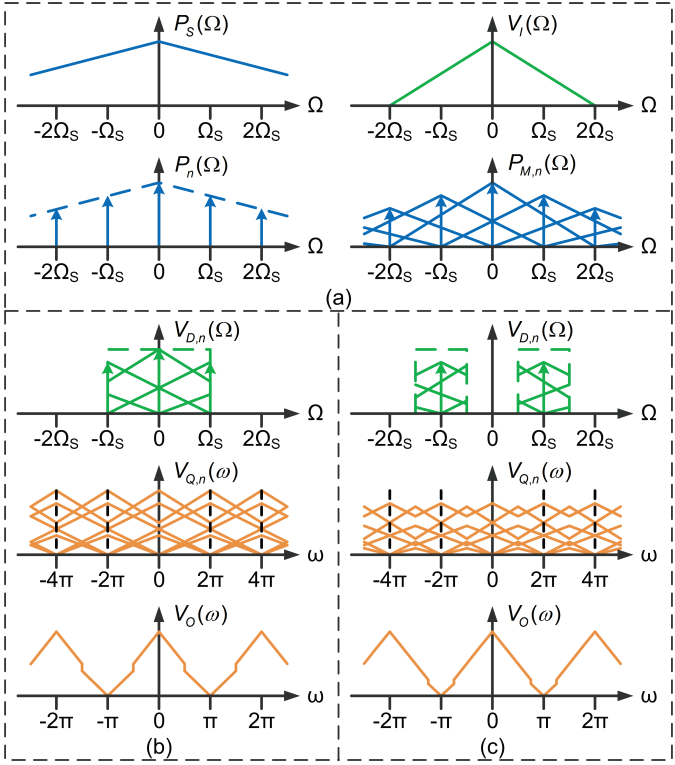

Fig. 2. Signal spectra of a 4-channel TIPADC in (a) the photonic sampling step, and the afterward steps with (b) lowpass $H_E(\Omega)$ and (c) bandpass $H_E(\Omega)$.

For typical applications, the following assumptions hold: 1) $P_S(\Omega)$ and $H_E(\Omega)$ are band-limited; 2) $P_S(\Omega)$ has a lowpass characteristic since the temporal shape of $p_S(t)$ is typically Gaussian-like; 3) $H_E(\Omega)$ has a lowpass or bandpass response. Under the above assumptions, we can determine the existence and the overlapping of the replicas of $H_E(\Omega)$ according to the bandwidth of $P_S(\Omega)$, the bandwidth and center frequency of $H_E(\Omega)$.

Since TIPADC aims to sample wideband signals, we will only consider the cases where the system bandwidth is not less than $f_S$. Then we have following constraints to the parameters (See Nomenclature for their definition):

$$R_{BP1}: \begin{cases} \chi_U - \chi_L \geq 1 \\ \beta_P > 0 \\ 2\chi_E \geq \beta_E \end{cases} \quad . \quad (24)$$

For description simplicity, we divide the replicas of $H_E(\Omega)$ into two parts: positive replicas and negative replicas, that are replicated from the non-negative frequency part and the non-positive frequency part of $H_E(\Omega)$, respectively. The non-negative frequency part of $H_E(\Omega)$ is the combination of the positive frequency part of $H_E(\Omega)$ and half of $H_E(0)$ at the zero frequency. Similarly the non-positive frequency part of $H_E(\Omega)$ is the combination of the negative frequency part of $H_E(\Omega)$ and half of $H_E(0)$. The $l$th positive replica and negative replica are centered at $\chi_E + l$ and $-\chi_E + l$, respectively. Since the weight $c_m$ of a replica outside the bandwidth of $P_S(\Omega)$ is zero, replicas exist only when the absolute value of the sequence number is not greater than $\Lambda_P$, the integer part of $\beta_P$.

The situations of the existence and the overlapping of replicas can be divided into 10 classes, as shown in Fig. 3. The inequalities in the figure are the constraints to the class located in corresponding row and column.

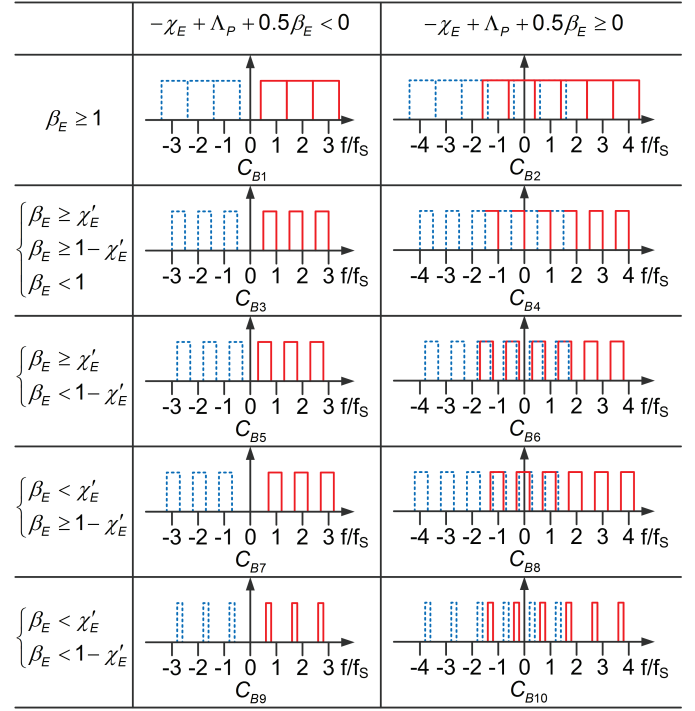

Fig. 3. 10 classes of the passband situation of $H_A(\Omega)$ for bandpass $H_E(\Omega)$. Red solid line: positive replicas, blue dashed line: negative replicas.

In the figure, $C_{Bi}$ ($i$=1, 3, 5, 7, 9) correspond to the cases where negative/positive replicas only exist at the negative/positive frequency axis, and $C_{Bj}$ ($j$=2, 4, 6, 8, 10) correspond to the cases where replicas can exist at the opposite frequency axis. Accordingly, the overlapping can only occurs among the same kind of replicas for $C_{Bi}$, while the overlapping can occur between both replicas for $C_{Bj}$. The largest system operating frequency range is determined by the lower and higher cutoff frequency of the passband that is formed by as many overlapped replica as possible. Therefore, all the ten classes can be divided into three situations: (i) overlapping occurs among at least three replicas, as in $C_{B1}$, $C_{B2}$ and $C_{B4}$. (ii) Overlapping occurs only between two replicas, as in $C_{B6}$ and $C_{B8}$. (iii) No overlapping occurs, as in $C_{B3}$, $C_{B5}$, $C_{B7}$, $C_{B9}$ and $C_{B10}$.

According to the above situations, the corresponding relationships between $\beta_E$ and the center frequency interval of replicas can be determined, that are listed in first column of Fig. 3. For $C_{B1}$ and $C_{B2}$, $\beta_E$ should be not less than 1 in order to guarantee the overlapping between the same kind replicas. $\beta_E$ for $C_{B3}$ and $C_{B4}$ should be larger than $Max(1-\chi'_E, \chi')$ and less than 1 at the same time, where $\chi'_E = 2\chi_E - \Lambda_E$ with $\Lambda_E = [2\chi_E]$. $\beta_E$ should be in the range of $[\chi'_E, 1-\chi'_E)$ for $C_{B5}$ to $C_{B6}$ and $[1-\chi'_E, \chi'_E)$ for $C_{B7}$ to $C_{B8}$, respectively. For $C_{B9}$ and $C_{B10}$, $\beta_E$ should be less than $Min(1-\chi'_E, \chi')$.

It can be directly derived that the bandwidths in the third situation are less than 1 since the replicas in these classes do not overlap with each other. In $C_{B6}$ and $C_{B8}$, the bandwidths can be expressed as,

$$\begin{aligned}\beta_{B6} &= \chi_E + (l - \Lambda_E) + 0.5\beta_E - (-\chi_E + l - 0.5\beta_E) \\ &= \chi'_E + \beta_E\end{aligned}, \quad (25)$$

$$\begin{aligned}\beta_{B8} &= -\chi_E + l + 0.5\beta_E - (\chi_E + l - \Lambda_E - 1 - 0.5\beta_E) \\ &= 1 - \chi'_E + \beta_E\end{aligned}, \quad (26)$$

respectively. Since $\beta_E$ is less than $1-\chi'_E$ or $\chi'_E$ in both classes, respectively, $\beta_{B6}$ and $\beta_{B8}$ are also less than 1. Considering the constraint of $R_{BP1}$, $C_{B3}$ and $C_{B5}$-$C_{B10}$ are not the feasible classes. Therefore we will not analyze them further. $C_{B1}$, $C_{B2}$ and $C_{B4}$ are the only classes whose bandwidth are likely to be larger than 1.

The lower and higher cutoff frequency of the passband in $C_{B1}$, $C_{B2}$ and $C_{B4}$ are shown in Table I. Combining the constraints to class $C_{B1}$ and $C_{B2}$, and the constraints to class $C_{B2}$ and $C_{B4}$, we have

$$R_{BP2}: \begin{cases} \beta_E \geq 1 \\ \chi_E - \Lambda_P - 0.5\beta_E \leq \chi_L \\ \chi_E + \Lambda_P + 0.5\beta_E \geq \chi_U \end{cases}, \quad (27)$$

$$R_{BP3}: \begin{cases} \beta_E \geq \chi'_E \\ \beta_E \geq 1 - \chi'_E \\ \chi_E + \Lambda_P - \Lambda_E + 0.5\beta_E \geq \chi_U \end{cases}. \quad (28)$$

The feasible region for TIPADCs can be expressed as:

$$R_{BP} = R_{BP4} \cup R_{BP5}, \quad (29)$$

where

$$R_{BP4} = R_{BP1} \cap R_{BP2} \text{ and } R_{BP5} = R_{BP1} \cap R_{BP3}. \quad (30)$$

The requirement to back-end electrical system in the feasible region can be derived by analyzing the minimum feasible $\beta_E$ as a function of other four parameters ($\beta_P$, $\chi_E$, $\chi_U$, $\chi_L$) in $R_{BP4}$ and $R_{BP5}$, respectively.

TABLE I
THE LOWER AND UPPER CUTOFF FREQUENCY OF THE PASSBAND HAVING THE LARGEST BANDWIDTH IN $C_{B1}$, $C_{B2}$ AND $C_{B4}$

| Class | The Lower Cutoff Frequency of The Passband | The Upper Cutoff Frequency of The Passband |
|---|---|---|
| $C_{B1}$ | the lower cutoff frequency of the $-\Lambda_P$ th positive replica | the upper cutoff frequency of the $\Lambda_P$ th positive replica |
| $C_{B2}$ | 0 | the upper cutoff frequency of the $\Lambda_P$ th positive replica |
| $C_{B4}$ | 0 | the upper cutoff frequency of the $(\Lambda_P - \Lambda_E)$ th positive replica |

The sub-region which contains all minimum feasible $\beta_E$ in $R_{BP4}$ can be derived as (detailed in Appendix I)

$$\begin{cases} 2\chi_E \geq \beta_E \geq 1 \\ \Lambda_P \geq \chi_E - 0.5 - \chi_L \\ \Lambda_P \geq \chi_H - \chi_E - 0.5 \\ \chi_U - \chi_L \geq 1 \end{cases}. \quad (31)$$

Equation (31) shows that the minimum feasible $\beta_E$ in $R_{BP4}$ is 1. It is reasonable and can be explained as follows. In the sub-region of (31), $\Lambda_P$ is high enough to make replicas exist in a wide frequency range where the formed passband can cover the designated system frequency range once adjacent replicas of the same kind overlap. Therefore, $\beta_E$ is only required to be greater than the center frequency interval of the adjacent replicas of the same kind, i.e., 1. In other sub-regions of $R_{BP4}$, $\Lambda_P$ is not high enough to keep replicas exist in a sufficient wide frequency range, and the minimum $\beta_E$ has to be greater than 1 to make the formed passband cover the designated system frequency range.

The sub-region which contains all minimum feasible $\beta_E$ in $R_{BP5}$ can be derived as (detailed in Appendix I)

$$\begin{cases} 2\chi_E \geq \beta_E \geq \chi'_E \\ \Lambda_P \geq \chi_U - \chi'_E + \frac{\Lambda_E}{2} \\ \chi'_E \geq 0.5 \\ \chi_U - \chi_L \geq 1 \end{cases} \cup \begin{cases} 2\chi_E \geq \beta_E \geq 1 - \chi'_E \\ \Lambda_P \geq \chi_U - \frac{1}{2} + \frac{\Lambda_E}{2} \\ 0.5 \geq \chi'_E \\ \chi_U - \chi_L \geq 1 \end{cases} \quad (32)$$

Similarly, in this sub-region, $\Lambda_P$ is high enough to make replicas exist in a wide frequency range, and $\beta_E$ should be large enough to ensure the overlapping between the adjacent replicas to form a passband covering the designated system frequency range. Since $\chi'_E$ varies with $\chi_E$, the minimum feasible $\beta_E$ changes periodically with the increasing of $\chi_E$. The global minimum value of $\beta_E$ is 0.5 when $\chi'_E$ is equal to 0.5. In this particular case, the sub-region containing all minimum feasible $\beta_E$ can be extracted from (32)

$$\begin{cases} 2\chi_E \geq \beta_E \geq 0.5 \\ \Lambda_P \geq \chi_U - 0.5 + 0.5\Lambda_E \\ \chi'_E = 0.5 \\ \chi_U - \chi_L \geq 1 \end{cases}. \quad (33)$$

For the case where $\Lambda_P$ is 0, i.e., $f_{B,P}$ is less than $f_S$, we can get by substituting it into $R_{BP4}$ and $R_{BP5}$,

$$\begin{cases} 2\chi_E \geq \beta_E \geq \chi_U - \chi_L \geq 1 \\ \chi_E - 0.5\beta_E \leq \chi_L \\ \chi_E + 0.5\beta_E \geq \chi_U \end{cases}. \quad (34)$$

We can see that the minimum feasible $\beta_E$ in this case is the designated system bandwidth $\chi_U - \chi_L$.

From above analyses, we can summarize the criteria in terms of frequency response for the TIPADCs whose bandwidth is greater than $f_S$ as follows.

1) The global minimum feasible $\beta_E$ is 0.5, if and only if $\beta_P$ and $\chi_E$ satisfy:

$$\begin{cases} \beta_P \geq \chi_U - 0.5 + 0.5\Lambda_E \\ \chi_E = 0.5k + 0.25, k \in \mathbb{N} \end{cases}; \quad (35)$$

2) The minimum feasible $\beta_E$ is not greater than 1, if and only if

$$\begin{cases} \beta_P \geq \chi_E - 0.5 - \chi_L \\ \beta_P \geq \chi_U - \chi_E - 0.5 \end{cases}; \quad (36)$$

3) The minimum feasible $\beta_E$ is not less than the designated system bandwidth $\chi_U - \chi_L$ when $\beta_P$ is less than 1.

Based on the analysis for bandpass $H_E(\Omega)$, the results for lowpass $H_E(\Omega)$ can be obtained by setting $\chi_E = 0.5\beta_E$. In this case, the situation of passband is degenerated into two classes, as shown in Fig. 4. $C_{L1}$ is degenerated from $C_{B2}$ and $C_{B4}$. $C_{L2}$ is degenerated from $C_{B6}$. The other classes in Fig. 3 cannot exist when $H_E(\Omega)$ is lowpass.

Similarly, only $C_{L1}$ is the feasible class under the constraint of $R_{BP1}$. The feasible region can be expressed as follows after some simple derivations.

$$R_{LP} : \begin{cases} \beta_E \geq 0.5 \\ \Lambda_P + \beta_E \geq \chi_U \\ \chi_U - \chi_L \geq 1 \end{cases}. \quad (37)$$

The feasible region can be further written as:

$$R_{LP} : \begin{cases} \beta_E \geq 0.5 \\ \Lambda_P \geq \chi_U - 0.5 \\ \chi_U - \chi_L \geq 1 \end{cases} \cup \begin{cases} \beta_E \geq \chi_U - \Lambda_P \\ \Lambda_P < \chi_U - 0.5 \\ \chi_U - \chi_L \geq 1 \end{cases}. \quad (38)$$

From (38), we can see that the global minimum feasible $\beta_E$ in $R_{LP}$ is 0.5, and the criteria for the TIPADC with a bandwidth not less than $f_S$ when $H_E(\Omega)$ is lowpass can be summarized as:

1) The global minimum feasible $\beta_E$ is 0.5 if $\beta_P$ satisfies

$$\beta_P \geq \chi_U - 0.5 \quad (39)$$

2) The minimum feasible $\beta_E$ is not less than the designated system bandwidth when $\beta_P$ is less than 1.

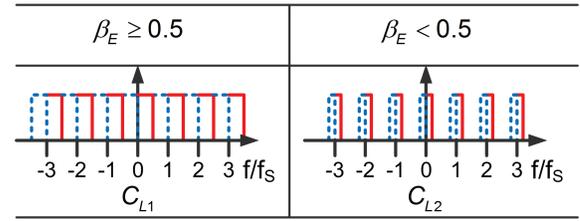

Fig. 4. 2 classes of the passband situation of $H_A(\Omega)$ for lowpass $H_E(\Omega)$. Red solid line: positive replicas, blue dashed line: negative replicas.

Fig. 5 illuminates the minimum feasible $\beta_E$ in $R_{BP}$ as a function of $\chi_E$ and $\beta_P$ for a lowpass or bandpass sampling system, respectively. From the figure, we can see that the global minimum feasible $\beta_E$ is 0.5, which is periodically reached as $\chi_E$ changes in the region indicated by criterion 1, just as the analysis to (32). When $\beta_P$ exceeds the threshold for criterion 2, the minimum feasible $\beta_E$ is always not greater than 1. The minimum feasible $\beta_E$ is not less than the system bandwidth (4 in the calculation) when $\beta_P$ is less than 1, as criterion 3 indicates.

*C. The Frequency Response of TIPADC under Different Conditions*

For a TIPADC satisfying the criteria, when $H_E(\Omega)$ complies with the Nyquist ISI criterion,

$$\sum_{m=-\infty}^{\infty} H_E(\Omega + m\Omega_S) = G_E \beta_E, \quad (40)$$

where $G_E$ is the gain in the passband of $H_E(\Omega)$, $H_A(\Omega)$ in the passband can be approximated as:

$$H_A(\Omega) \approx 2a_1 P_A R_D R_L G_E \beta_E. \quad (41)$$

The constant value implies that there is no ripple in passband. Conversely when $H_E(\Omega)$ does not satisfy the Nyquist ISI, there may be significant ripple in the passband of $H_A(\Omega)$ since the superposition of the replicas of $H_E(\Omega)$ is no more a constant as (40) indicates. Fig. 6 shows the theoretical $H_A(\Omega)$ (Note that $f$ rather than $\Omega$ is used in the related figures below for conventional compliance) and its approximation for different $\beta_P$ and $H_E(\Omega)$ of $\eta$-order lowpass Butterworth filter. The other parameters are $P_A$ of 3dBm, $V_\pi$ of 3.14V, $R_D$ of 1A/W, $R_L$ of 50 $\Omega$, $G_E$ of 20dB. Because $H_E(\Omega)$ only approximately satisfies (40) for system configurations shown in Fig. 6, there is ripple in the passband. Nevertheless, the approximation is still coincident in both cases.

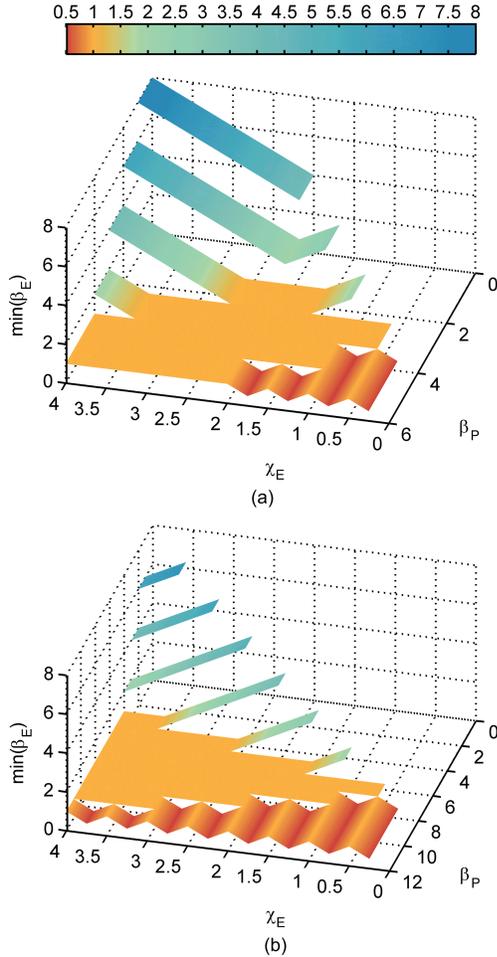

Fig. 5. The minimum feasible $\beta_E$ as a function of $\chi_E$ and $\beta_P$ when $\chi_L$ and $\chi_U$ are: (a) 0 and 4, (b) 4 and 8.

Based on the foregoing analysis, we explore $H_A(\Omega)$ for a 32-channel TIPADC with $f_S$ of 1Gs/s. Fig. 7 shows $H_A(\Omega)$ with different $H_E(\Omega)$ when $\beta_P$ is 16.5. The other parameters are the same as those in Fig. 6. In Fig. 7(a), $H_E(\Omega)$ is bandpass with $\chi_E$ of 0.75 and $\eta$ of 4. We can see that $\beta_E$ with value of 1 or 0.5 is enough for sampling, while $\beta_E = 0.25$ is not. This is because the system parameters satisfy the conditions of criterion 1 for bandpass $H_E(\Omega)$. The results validate criterion 1 that states the minimum feasible $\beta_E$ is 0.5 under the specified conditions. In Fig. 7(b), the $\chi_E$ of bandpass $H_E(\Omega)$ increases to 1.5, which leads to the system parameters can only satisfy criterion 2, rather than criterion 1. Accordingly, we can see that the minimum feasible $\beta_E$ is increased to 1. In Fig. 7(c), $H_E(\Omega)$ is lowpass with $\eta$ of 2. The system parameters satisfy criterion 1. The result is similar to those shown in Fig. 7(a) and consistent with the theoretical analysis. The ripples in the passband in all three cases are attributed to that $H_E(\Omega)$ does not satisfy ISI criterion, as mentioned in the derivation of (41).

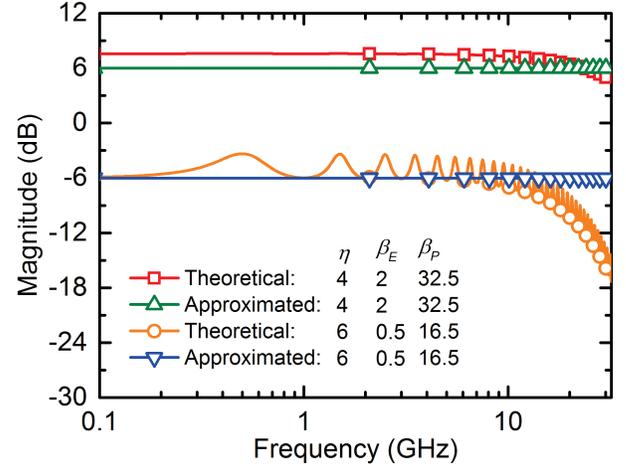

Fig. 6. Theoretical and approximated $H_A(\Omega)$ for different $H_E(\Omega)$ and $\beta_P$.

Fig. 8 shows $H_A(\Omega)$ for different $\beta_P$ when $H_E(\Omega)$ is bandpass with $\eta$ of 4, $\beta_E$ of 0.5, and $\chi_E$ of 0.75, or lowpass with $\eta$ of 2 and $\beta_E$ of 0.5, respectively. The other parameters are the same as in Fig. 6. One can see that, no matter weather $H_E(\Omega)$ is lowpass or bandpass, the TIPADC is capable to sample the signals with frequency up to 8GHz and 16GHz for $\beta_P$ of 8.5 and 16.5, respectively. The result is consistent with criterion 1. When $\beta_P$ approaches 1, the system frequency response approaches $H_E(\Omega)$, which is consistent with criterion 3.

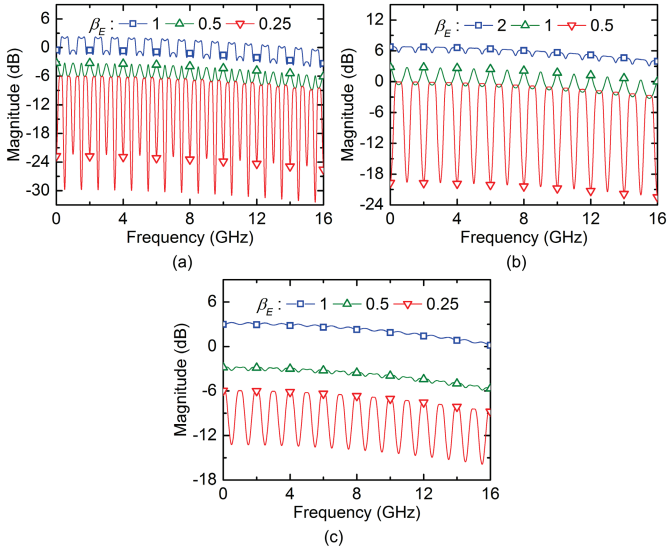

Fig. 7. System frequency responses for different $H_E(\Omega)$: bandpass in (a) and (b), and lowpass in (c).

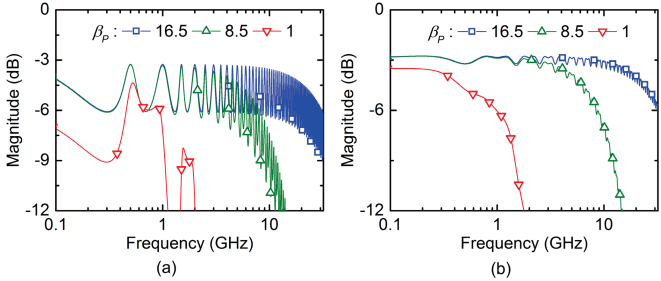

Fig. 8. System frequency response under different $\beta_P$ for $H_E$ of (a) bandpass and (b) lowpass.

## IV. THE NOISE AND NONLINEAR PERFORMANCE OF TIPADC

In this section, we derive the output powers of different noises and nonlinear distortion from the model, and analyze the effect of different factors on system ENOB. For simplicity, we assume that $v_{M,n}(t) = V_M \cos(\Omega_I t + \theta_{M,n})$.

### A. Amplitude Noise and Phase Noise in the Photonic Sampling Pulse Train

The generated photonic sampling pulse train usually fluctuates randomly both in amplitude and phase. If they are not related, we can characterize them by the amplitude fluctuation, $\Delta P_{A,n}$, and the timing jitter, $\Delta t_{P,n}$, and analyze them independently. If $\Delta t_{P,n} \ll T_S$, as is usually required in the practical implementation of TIPADC, the noise of photonic sampling pulse train can be expressed as [18]-[19]

$$\Delta p_n(t) \approx P_A \sum_{m=-\infty}^{\infty} c_m e^{jm\Omega_S t} \left[ \Delta_{A,n}(t) + j2\pi m \Delta_{P,n}(t) \right], \quad (42)$$

where the relative amplitude fluctuation $\Delta_{A,n} = \Delta P_{A,n}/P_A$ and relative timing jitter $\Delta_{P,n} = \Delta t_{P,n}/T_S$ are assumed to be zero-mean wide-sense stationary processes. Also their powers, $\sigma_{A,n}^2$ and $\sigma_{P,n}^2$, are assumed to be equal for all channels, i.e., $\sigma_{A,n}^2 = \sigma_A^2$ and $\sigma_{P,n}^2 = \sigma_P^2$, respectively.

In the presence of only the amplitude noise, the noise power of system output can be approximated as (detailed in Appendix II)

$$\sigma_O^2 \approx \left[ \frac{a_0^2}{a_1^2} |H_A(0)|^2 + \frac{V_M^2}{2} |H_A(\Omega_I)|^2 \right] \sigma_A^2. \quad (43)$$

When the approximation for $H_A(\Omega)$ is valid, it can be further approximated as

$$\sigma_O^2 \approx (P_A R_D R_L G_E \beta_E \sigma_A)^2 \left( 4a_0^2 + 2a_1^2 V_M^2 \right). \quad (44)$$

SNR in the presence of only the amplitude noise can be approximated as

$$SNR_A = \frac{V_M^2 |H_A(\Omega_I)|^2}{2\sigma_O^2} \approx \frac{1}{\left( 2a_0^2 / a_1^2 V_M^2 + 1 \right) \sigma_A^2}. \quad (45)$$

It indicates that the SNR in the presence of only the amplitude noise is related to the modulation depth of EOM, $a_1 V_M / a_0$, and $\sigma_A$. The SNR cannot be improved by adjusting other parameters and is uniform to all input frequency. This is a fundamental performance limitation induced by the photonic sampling pulse train.

Similarly, in the presence of only the phase noise, the noise power of system output can be approximated as (detailed in Appendix II)

$$\sigma_O^2 \approx \left[ \frac{a_0^2}{a_1^2} |H_P(0)|^2 + \frac{V_M^2}{2} |H_P(\Omega_I)|^2 \right] \sigma_P^2, \quad (46)$$

where $H_P(\Omega)$ is the frequency response of system to the phase noise and is given by

$$H_P(\Omega) = j2\pi a_1 P_A R_D R_L \sum_{m=-\infty}^{\infty} m c_m H_E(\Omega + m\Omega_S). \quad (47)$$

In the same condition as deriving the approximation of $H_A(\Omega)$, there is an approximation for $H_P(\Omega)$ (detailed in Appendix II):

$$H_P(\Omega) \approx \frac{j4\pi a_1 P_A R_D R_L G_E \beta_E \Omega}{\Omega_S}. \quad (48)$$

Fig. 9 shows theoretical $H_P(\Omega)$ and its approximation for the same parameters as in Fig. 6.

When (48) is valid, we can further approximate $\sigma_O^2$ as

$$\sigma_O^2 \approx 8 \left( \pi a_1 P_A R_D R_L G_E \chi_I \beta_E V_M \sigma_P \right)^2, \quad (49)$$

where $\chi_I = \Omega_I / \Omega_S$.

Hence in the presence of only the phase noise, SNR can be approximated as

$$SNR_P \approx \frac{1}{4\pi^2 \chi_I^2 \sigma_P^2}. \tag{50}$$

The formula explicitly derived in this model confirms the conclusion about the phase noise of the photonic sampling pulse train in [2]. We can see that the SNR will decrease with the increase of input frequency.

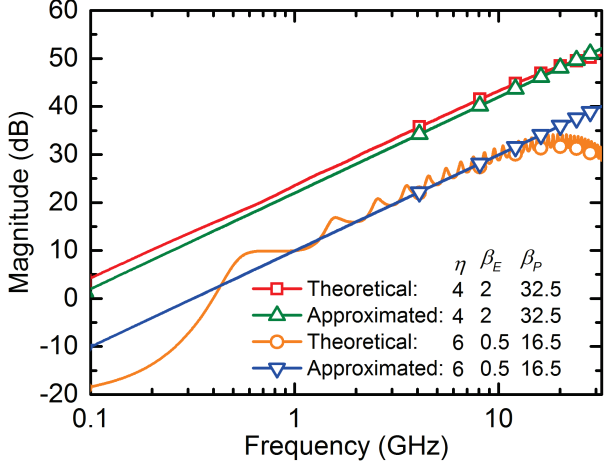

Fig. 9. Theoretical $H_P(\Omega)$ and its approximation for different $H_E(\Omega)$ and $\beta_P$.

### B. Thermal Noise and Shot Noise in Photodetection

Thermal noise originates from the thermal vibration of the electrons, which can be characterized by the noise temperature $T_E$. Assuming both the noise temperature $T_E$ are the same, the noise power of system output in the presence of only thermal noise can be written, no matter $H_E(\Omega)$ is lowpass or bandpass, as

$$\sigma_O^2 = KT_E R_L G_D^2 \beta_E f_S \tag{51}$$

where $K$ is Boltzmann's constant. Hence the SNR in this case can be derived as

$$SNR_T = \frac{2\beta_E R_L (a_1 P_A R_D V_M)^2}{KT_E f_S}. \tag{52}$$

Shot noise originates from the quantum nature of the photons in photodetection. Since the shot noise power is proportional to its mean [14]-[15], it can be characterized by $P_A$ when $v_{M,n}(t)$ is cosine wave. The power of system output in the presence of only shot noise can be written as

$$\sigma_O^2 = 2qa_0 P_A R_D R_L^2 G_D^2 \beta_E f_S, \tag{53}$$

where $q$ is the fundamental charge. Hence the SNR in this case can be derived as

$$SNR_S = \frac{\beta_E R_D P_A a_1^2 V_M^2}{qa_0 f_S}. \tag{54}$$

### C. Quantization Noise and Aperture jitter Noise in EADCs

When EADCs digitize the photodetected signal, there are quantization noise and aperture jitter noise. Quantization noise originates from the amplitude discretizing to the input of EADCs. As a well-known result, when $v_{M,n}(t)$ is cosine wave and the voltage range of the photodetected signal does not exceed the input span of EADC, the noise power of system output in the presence of only quantization noise can be written as

$$\sigma_O^2 = \frac{V_A^2}{12 \times 4^{N_A}}, \tag{55}$$

where $N_A$ is the resolution of EADCs, $V_A$ is the voltage input span. The SNR in this case can be derived as

$$SNR_Q = 24 \times 4^{N_A} \left( \frac{a_1 P_A R_D R_L G_E \beta_E V_M}{V_A} \right)^2. \tag{56}$$

The noise generated from the aperture jitter of EADC is related to the derivative of $v_{D,n}(t)$ and the relative jitter $\Delta_{J,n} = \Delta t_{J,n}/T_S$, and can be derived as follows under the condition $\Delta_{J,n} \ll 1$:

$$\Delta v_{Q,n}[k] \approx T_S v'_{D,n}(t) \Delta_{J,n}(t) \tag{57}$$

Similarly, in the presence of only the aperture jitter noise, the noise power of system output can be obtained as (detailed in Appendix II)

$$\sigma_O^2 \approx \left[ \frac{a_0^2}{a_1^2} |H_J(0)|^2 + \frac{V_M^2}{2} |H_J(\Omega_I)|^2 \right] \sigma_J^2, \tag{58}$$

where $H_J(\Omega)$ is the frequency response of system to the aperture jitter noise and is given by

$$H_J(\Omega) = ja_1 P_A R_D R_L T_S$$
$$\cdot \sum_{m=-\infty}^{\infty} c_m (\Omega + m\Omega_S) H_E(\Omega + m\Omega_S). \tag{59}$$

Note that there is a relationship among $H_A(\Omega)$, $H_P(\Omega)$ and $H_J(\Omega)$ according to (22), (47) and (59):

$$j\Omega T_S H_A(\Omega) = H_P(\Omega) + H_J(\Omega). \tag{60}$$

Under the same condition as those in deriving the approximation of $H_A(\Omega)$, we have the approximation for $H_J(\Omega)$ from (60):

$$H_J(\Omega) \approx 0. \tag{61}$$

It implies that the requirement to aperture jitter of EADCs can be considerably alleviated compared to the jitter in the photonic sampling pulse train.

Fig. 10 shows theoretical $H_J(\Omega)$ for the same parameters as in Fig. 6. One can see that $H_J(\Omega)$ is very small in comparison with $H_P(\Omega)$ shown in Fig. 9 and is not proportional to $\chi_I$ on the trend.

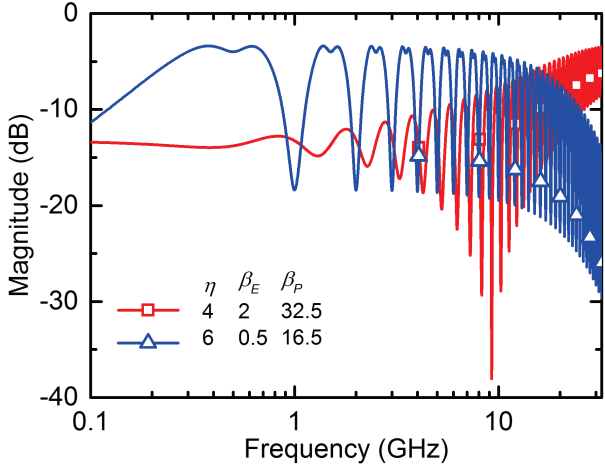

Fig. 10. Theoretical $H_J(\Omega)$ for different $H_E(\Omega)$ and $\beta_P$.

### D. Nonlinearity in TIPADC

In TIPADC, nonlinear distortion mainly arises from the nonlinear response of the modulator, $t_M(v)$, when $P_A$ is moderate and within the linear input power range of photodetector (so that nonlinearity of photodetector can be ignored) [14]-[15]. In the above frequency response analysis, $t_M(v)$ is linearized under the small signal approximation to obtain the linear characteristics. Next, we characterize the nonlinear performance for quadrature-biased Mach-Zehnder modulator. When the input RF signal is a single tone, according to Jacobi–Anger expansion, we have

$$t(v_{M,n}) = \frac{1}{2} + \sum_{m=1}^{\infty} (-1)^m J_{2m-1}\left(\frac{\pi V_M}{V_\pi}\right) \cdot \cos\left[(2m-1)\Omega_I t + (2m-1)\theta_{M,n}\right], \quad (62)$$

where $J_m(x)$ is the first kind of Bessel function of order $m$. (62) implies that the fundamental and harmonics are linear superposition. Hence all the nonlinear components can be referred to input and system is still linear. We can obtain the harmonics power of system output from the model in Section III

$$\sigma_O^2 = \frac{1}{2a_1^2}\sum_{m=2}^{\infty} J_{2m-1}^2\left(\frac{\pi V_M}{V_\pi}\right)\left|H_A(2m\Omega_I - \Omega_I)\right|^2. \quad (63)$$

We only consider the third harmonic since it is the dominant in typical cases. Then according to the definition of the signal to distortion ratio (SDR), we have

$$SDR \approx \frac{J_1^2(\pi V_M/V_\pi)|H_A(\Omega_I)|^2}{J_3^2(\pi V_M/V_\pi)|H_A(3\Omega_I)|^2}. \quad (64)$$

### E. ENOB Considerations

ENOB is related to the signal to noise and distortion ratio (SINAD) of converters and is measured under single-tone stimulus [20]:

$$ENOB = \frac{SINAD - 1.76}{6.02}. \quad (65)$$

Table II summaries the influence of various factors on different SNRs and SDR when the approximation for $H_A(\Omega)$ is valid. From the table, the following deductions can be made.

TABLE II
THE TREND OF DIFFERENT SNRs AND SDR AS THE INCREASING OF VARIOUS FACTORS IN THE MODEL WHEN THE APPROXIMATION FOR $H_A(\Omega)$ IS VALID

|         | $N$ | $f_S$ | $P_A$ | $R_D$ | $R_L$ | $\beta_E$ | $G_E$ | $\frac{V_M}{V_\pi}$ |
|---------|-----|-------|-------|-------|-------|-----------|-------|---------------------|
| $SNR_A$ | –   | –     | –     | –     | –     | –         | –     | ↑                   |
| $SNR_P$ | –   | –     | –     | –     | –     | –         | –     | –                   |
| $SNR_T$ | –   | ↓     | ↑     | ↑     | ↑     | ↑         | –     | ↑                   |
| $SNR_S$ | –   | ↓     | ↑     | ↑     | –     | ↑         | –     | ↑                   |
| $SNR_Q$ | –   | –     | ↑     | ↑     | ↑     | ↑         | ↑     | ↑                   |
| $SNR_J$ | –   | –     | –     | –     | –     | –         | –     | –                   |
| $SDR$   | –   | –     | –     | –     | –     | –         | –     | ↓                   |

The number of channels, $N$, is not related to ENOB under the above assumption, which indicates that increasing total sampling rate by increasing $N$ will not degrade the ENOB performance in ideal case. In practical, however channel mismatch is a primary issue for multichannel TIPADC. It may significantly degrade ENOB by introducing mismatch spurs [21]-[22].

$f_S$ is inversely proportional to SNR in the presence of thermal and shot noise. This is because increasing $f_S$ implies that the bandwidth of back-end electronics should be increased for a given $\beta_E$ and thus the thermal and shot noise powers are increased. Therefore, there is a trade-off between ENOB and sampling rate.

As $P_A$ increases, $SNR_T$, $SNR_S$ and $SNR_Q$ are improved. This improvement, however, is limited by the linear input range of photodetector, and the input span of EADCs. Moreover, when $P_A$ is high enough, the total SNR will be dominated by the noises whose SNR is not related to $P_A$ and no further improvement can be imposed by increasing $P_A$.

$\beta_E$ is limited by the above criteria. Lower $\beta_E$ allows the use of EADCs with lower bandwidth and higher ENOB, which reduces the difficulty in implementing back-end electronics and supports system operation with higher ENOB. On the other hand, higher $\beta_E$ is preferred in terms of thermal noise, shot noise, and quantization noise from Table II. Therefore, an optimization for $\beta_E$ is needed in the system design.

$G_E$ should be sufficient to minimize the penalty induced by

quantization noise. Quantization noise sets a basic limitation to the system resolution, leading to an ENOB lower than $N_A$. Only when a single-tone input signal reaches full-scale, ENOB is possible to reach $N_A$.

For SNR and SDR, the requirement on $V_M/V_\pi$ is contrary to each other. The former prefers a higher $V_M/V_\pi$, while the later prefers lower $V_M/V_\pi$. Fig. 11 shows theoretical ENOB versus $V_M/V_\pi$ in the presence of nonlinear distortion and different noises, respectively. In the figure, $H_E(\Omega)$ is lowpass with $\eta$ of 2, $\beta_E$ of 0.5 and $G_E$ of 40dB; $f_S$ =1Gs/s; $\beta_P$ =100.5; $f_I$ =1GHz; $P_A$ =10dBm; $V_\pi$ =3.14V; $R_D$ =1A/W; $R_L$ =50Ω; $\sigma_A$ =-80dB; $\sigma_P$ =-114dB (2fs jitter); $\sigma_J$ =-74dB (200fs jitter); and $T_E$ =300K. As the figure shows, ENOB in the presence of only the thermal noise or shot noise is proportional to $V_M/V_\pi$. ENOBs in the presence of only the amplitude noise, phase noise, or aperture jitter noise trend to be saturated as $V_M/V_\pi$ increases. The reason can be explained as follows. As $V_M/V_\pi$ increases, the dominant component in the noise trend to change from the unmodulated component to the modulated one. The noise power related to unmodulated component is not related to $V_M/V_\pi$ while the noise power related to the modulated one is proportional to the square of $V_M/V_\pi$. The output signal power is also proportional to the square of $V_M/V_\pi$. Therefore, ENOBs in the presence of these noises trend to saturation as the increase of $V_M/V_\pi$.

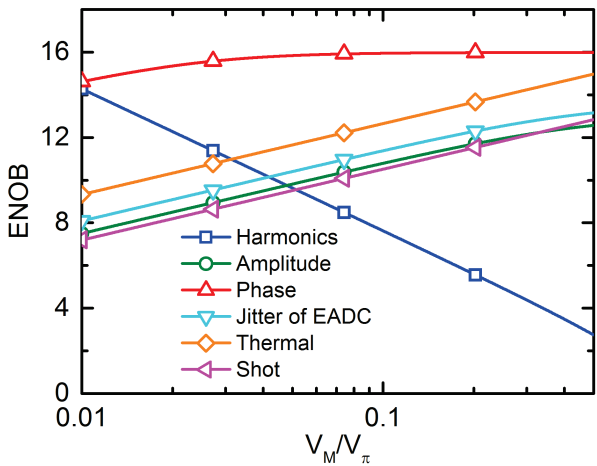

Fig. 11. Theoretical ENOB versus $V_M/V_\pi$ in the presence of nonlinear distortion and various noises.

Fig. 12 shows the theoretical ENOB of a TIPADC versus input frequency in the presence of nonlinear distortion and different noises when $V_M/V_\pi = 0.05$ and other parameters are the same as those in Fig. 11. Note that we ignore the limited bandwidth of EOM, i.e., $H_M(\Omega)$ =1. In fact, EOM with bandwidth over 100GHz is now commercially available [23]. From the figure, the TIPADC has about 100GHz bandwidth, which is consistent with the criteria. The ENOB related to the phase noise always gets lower as $f_I$ increases. The ENOB related to the harmonics shows a significant increasing trend when $f_I$ is beyond almost 30GHz. It is due to that when $f_I$ is high enough the harmonics lie outside the system passband. However, for non-single-tone input, nonlinear components may lie inside the system passband and the effect of nonlinearity is still severe. To remove this ambiguity, the nonlinearity performance for all $f_I$ can refer to that in low $f_I$.

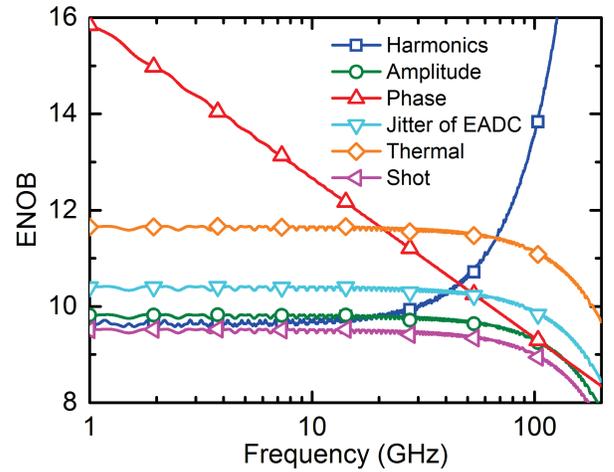

Fig. 12. Theoretical ENOB versus $f_I$ in the presence of nonlinear distortion and various noises.

## V. CONCLUSION

We proposed a theoretical model for the TIPADC to describe its operation principle, including photonic sampling and demultiplexing, photodetection, electronic quantization and digital processing. Based on the model, the signal sampling mechanism of TIPADCs was illustrated, and the sampling output and frequency response of system in ideal conditions were derived. We revealed that photonic sampling folds the whole spectrum of input signal into a narrow band, which enables the analog bandwidth of a TIPADC to be much higher than the bandwidth of back-end electronics. In terms of frequency response, the feasible region of system parameters and the sampling criteria of TIPADC for the typical applications were presented and the global minimum feasible bandwidths of back-end electronics were given. The criteria set a primary requirement to the system performance and are effective supplement to the generalized sampling theorem for TIPADC systems. The analysis based on the system frequency response showed that the ISI induced by the limited bandwidth of back-end electronics will only cause ripple in passband. The major noises and nonlinearity were investigated to explore the effects of different parameters to system ENOB. The system ENOB was characterized and its bottleneck in different cases was figured out. This study is concentrated on the case that sampling channels are matched and gives the basic performance description. The mismatches and crosstalk between sampling channels are worth investigating and will be our further work.

## APPENDIX I

This appendix details the derivation of (31) and (32).

$R_{BP4}$ can be expressed as

$$\begin{cases} \beta_E \geq 1 \\ \beta_E \geq 2\chi_E - 2\Lambda_P - 2\chi_L \\ \beta_E \geq 2\chi_U - 2\chi_E - 2\Lambda_P \\ 2\chi_E \geq \beta_E \\ \chi_U - \chi_L \geq 1 \end{cases}. \quad (A.1)$$

From (A.1), we can see that there are three lower limits on $\beta_E$ in $R_{BP4}$: 1, $(2\chi_E - 2\Lambda_P - 2\chi_L)$, and $(2\chi_U - 2\chi_E - 2\Lambda_P)$. For a certain set of parameter values of $\chi_E$, $\Lambda_P$, $\chi_U$ and $\chi_L$, the minimum feasible $\beta_E$ in this case is the largest one among the three limits. Therefore, in $R_{BP4}$ the sub-region where the minimum feasible $\beta_E$ is 1 should satisfy following conditions:

$$\begin{cases} 2\chi_E - 2\Lambda_P - 2\chi_L \leq 1 \\ 2\chi_U - 2\chi_E - 2\Lambda_P \leq 1 \end{cases}. \quad (A.2)$$

Combining (A.1) and (A.2), we can obtain (31). It is not a null set. At the same time, since in $R_{BP4}$ the complement of (31) is the same as that of (A.2), the minimum feasible $\beta_E$ in the complement region of (31) is greater than 1. That means the minimum feasible $\beta_E$ in $R_{BP4}$ is 1 and all the point with the minimum feasible $\beta_E$ are contained in (31).

$R_{BP5}$ can be expressed as

$$\begin{cases} \beta_E \geq \chi'_E \\ \beta_E \geq 1 - \chi'_E \\ \beta_E \geq 2\chi_U - 2\chi_E - 2(\Lambda_P - \Lambda_E) \\ 2\chi_E \geq \beta_E \\ \chi_U - \chi_L \geq 1 \end{cases}. \quad (A.3)$$

From (A.3), we can also see that there are three lower limits on $\beta_E$ in $R_{BP5}$: $\chi'_E$, $1-\chi'_E$, and $2\chi_U - 2\chi_E - 2(\Lambda_P - \Lambda_E)$. Since the range of $\chi'_E$ is $[0,1)$, we can find that if $\chi'_E$ is not less than 0.5, the minimum feasible $\beta_E$ should be not less than $\chi'_E$, and if $\chi'_E$ is not greater than 0.5, the minimum feasible $\beta_E$ should be not less than $1 - \chi'_E$. Therefore the sub-region, where the minimum feasible $\beta_E$ is $\chi'_E$, should satisfy

$$\begin{cases} \chi'_E \geq 0.5 \\ 2\chi_U - 2\chi_E - 2(\Lambda_P - \Lambda_E) \leq \chi'_E \end{cases}, \quad (A.4)$$

and the sub-region, where the minimum feasible $\beta_E$ is $1 - \chi'_E$, should satisfy

$$\begin{cases} \chi'_E \leq 0.5 \\ 2\chi_U - 2\chi_E - 2(\Lambda_P - \Lambda_E) \leq 1 - \chi'_E \end{cases}. \quad (A.5)$$

Combining (A.3) with (A.4) and (A.5), we obtain (32). Similarly, we can see that the minimum feasible $\beta_E$ in $R_{BP5}$ is 0.5 and all of them are contained in (32).

## APPENDIX II

This appendix gives the detailed derivation of the noise powers of system output.

First we deal with the noises in the photonic sampling pulse train. According to the modelling in Section II, we can derive the noise in samples that is digitized by EADCs:

$$\Delta v_{Q,n}[k] = R_D R_L h_E(t) \\ * [a_0 \Delta p_n(t) + a_1 v_{M,n}(t) \Delta p_n(t)]\big|_{t=kT_S}. \quad (A.6)$$

Under the assumption that the amplitude and the phase noise are not related, we can analyze them independently. Utilizing the periodicity of $p_n(t)$, we can derive the autocorrelation of $\Delta v_{Q,n}[k]$ in the present of only the amplitude noise:

$$\begin{aligned} R_{\Delta Q,n}[l] &= \lim_{K \to \infty} \frac{1}{K} \sum_{k=0}^{K-1} E\big[\Delta v_{Q,n}[k+l] \Delta v_{Q,n}[k]\big] \\ &= R_D^2 R_L^2 \lim_{K \to \infty} \frac{1}{K} \sum_{k=0}^{K-1} \iint \big[a_0 \\ &\quad + a_1 V_M \cos(k\chi_I - \Omega_I \tau_2 + \theta_{M,n})\big] \\ &\quad \cdot \big[a_1 V_I \cos(k\chi_I + l\chi_I - \Omega_I \tau_1 + \theta_{M,n}) \\ &\quad + a_0\big] p_n(-\tau_2) p_n(-\tau_1) h_E(\tau_2) h_E(\tau_1) \\ &\quad \cdot R_A(lT_S - \tau_1 + \tau_2) d\tau_2 d\tau_1 \end{aligned}. \quad (A.7)$$

Exchanging the order of limit and integral, we can obtain the result of time averaging:

$$\begin{aligned} R_{\Delta Q,n}[l] &= R_D^2 R_L^2 \\ &\quad \cdot \iint \left[a_0^2 + \frac{a_1^2 V_M^2}{2} \cos(l\chi_I - \Omega_I \tau_1 + \Omega_I \tau_2)\right] \\ &\quad \cdot p_n(-\tau_2) p_n(-\tau_1) h_E(\tau_2) h_E(\tau_1) \\ &\quad \cdot R_A(lT_S - \tau_1 + \tau_2) d\tau_2 d\tau_1 \end{aligned}. \quad (A.8)$$

Simplifying (A.8), we can derive the autocorrelation of $\Delta v_{Q,n}[k]$ in the presence of only the amplitude noise:

$$R_{\Delta Q,n}[l] = \lim_{K \to \infty} \frac{1}{K} \sum_{k=0}^{K-1} E\left[\Delta v_{Q,n}[k+l]\Delta v_{Q,n}[k]\right]$$
$$= \left[\frac{a_0^2}{a_1^2} + \frac{V_M^2}{2}\cos(\Omega_I \tau)\right] R_{A,n}(\tau) \quad \text{(A.9)}$$
$$\left. * h_A(\tau) * h_A(-\tau)\right|_{\tau=lT_S}$$

where the autocorrelation of $\Delta_{A,n}$ is denoted by $R_{A,n}(\tau)$.

Then we can derive the noise power of samples in the presence of only the amplitude noise from the power spectrum density (PSD), $S_{\Delta Q,n}(w)$, the Fourier transform of $R_{\Delta Q,n}[l]$, according to Wiener-Khinchin theorem:

$$\sigma_{Q,n}^2 = \frac{1}{2\pi}\int_{-\pi}^{\pi} S_{\Delta Q,n}(w) d\omega$$
$$= \frac{1}{2\pi}\int_{-\infty}^{\infty} |H_A(\Omega)|^2 \left[\frac{V_M^2}{4} S_{A,n}(\Omega + \Omega_I)\right. \quad \text{(A.8)}$$
$$\left. + \frac{a_0^2}{a_1^2} S_{A,n}(\Omega) + \frac{V_M^2}{4} S_{A,n}(\Omega - \Omega_I)\right] d\Omega$$

where $S_{A,n}(\Omega)$ is the PSD of $\Delta_{A,n}$.

We can derive an approximation for $\sigma_{Q,n}^2$. In its expression, the term related to $S_{A,n}(\Omega)$ arises from the amplitude noise on the original photonic sampling pulse train and the terms related to $S_{A,n}(\Omega + \Omega_I)$ and $S_{A,n}(\Omega - \Omega_I)$ arise from the amplitude noise on the modulated photonic sampling pulse train. $H_A(\Omega)$ is the frequency response of the system to the amplitude noise. We approximate the response of $S_{A,n}(\Omega)$ to be $H_A(0)$ since the power of $S_{A,n}(\Omega)$ is concentrate on low frequency. Similarly, the responses to $S_{A,n}(\Omega + \Omega_I)$ and $S_{A,n}(\Omega - \Omega_I)$ are approximated to $H_A(\Omega_I)$ and $H_A(-\Omega_I)$, respectively. Hence we have the approximation (43).

Under the case where $H_A(\Omega)$ can be approximated by (41), the noise power of samples can be derived as

$$\sigma_{Q,n}^2 \approx \left(P_A R_D R_L G_E \beta_E \sigma_A\right)^2 \left(4a_0^2 + 2a_1^2 V_M^2\right) \quad \text{(A.10)}$$

Considering the samples of all channels are interleaved together, we can obtain the relation between the noise power of system output $\sigma_O^2$ and $\sigma_{Q,n}^2$,

$$\sigma_O^2 = \lim_{K \to \infty} \frac{1}{KN} \sum_{k=0}^{K-1}\sum_{n=0}^{N-1} E\left[\Delta v_O^2[kN+n]\right]$$
$$= \lim_{K \to \infty} \frac{1}{KN} \sum_{k=0}^{K-1}\sum_{n=0}^{N-1} E\left[\Delta v_{Q,n}^2[k]\right] \quad \text{(A.11)}$$
$$= \frac{1}{N}\sum_{n=0}^{N-1}\sigma_{Q,n}^2$$

Hence from (A.9) and (A.11), (43) is obtained. Substituting the approximation for $H_A(\Omega)$ in (43), (44) is derived.

Similarly, for the phase noise in photonic sampling pulse train, we can derive the autocorrelation of $\Delta v_{Q,n}[k]$ from the model:

$$R_{\Delta Q,n}[l] = \left[\frac{a_0^2}{a_1^2} + \frac{V_M^2}{2}\cos(\Omega_I \tau)\right] R_{P,n}(\tau) \quad \text{(A.12)}$$
$$\left. * h_P(\tau) * h_P(-\tau)\right|_{\tau=lT_S}$$

where $h_P(t)$ is the impulse response of system to the phase noise

$$h_P(t) = a_1 R_D R_L T_S p_n'(-t) h_E(t), \quad \text{(A.13)}$$

and $p_n'(t)$ is the derivative of $p_n(t)$.

In the presence of only the phase noise, we can obtain the noise power in samples similarly:

$$\sigma_{Q,n}^2 = \frac{1}{2\pi}\int_{-\infty}^{\infty} |H_P(\Omega)|^2 \left[\frac{V_M^2}{4} S_{P,n}(\Omega + \Omega_I)\right.$$
$$\left. + \frac{a_0^2}{a_1^2} S_{P,n}(\Omega) + \frac{V_M^2}{4} S_{P,n}(\Omega - \Omega_I)\right] d\Omega \quad \text{(A.14)}$$

where $S_{P,n}(\Omega)$ is the PSD of $\Delta_{P,n}$.

Considering the power of $S_{P,n}(\Omega)$ is also concentrate on low frequency, we can obtain the approximation for $\sigma_{Q,n}^2$ in the presence of only the phase noise similarly and have (46).

An approximation for $H_P(\Omega)$ can be derived similar to $H_A(\Omega)$. Under the case where $H_A(\Omega)$ can be approximated by (41), $H_P(\Omega)$ at a given frequency is contributed by several replicas whose weights are arithmetic series approximately. Then $H_P(\Omega)$ can be approximated to the sum of arithmetic series whose mean and number of terms are about $\Omega/\Omega_S$ and $2\beta_E$, respectively. Hence in the passband of $H_A(\Omega)$, we have (48).

In the presence of only the aperture jitter noise, from (57), the noise in samples can be expressed as

$$\Delta v_{Q,n}[k] \approx T_S R_D R_L \Delta_{J,n}(t) \left\{h_E'(t) * [a_0 \Delta p_n(t)\right.$$
$$\left. + a_1 v_{M,n}(t)\Delta p_n(t)]\right\}\Big|_{t=kT_S} \quad \text{(A.15)}$$

Then, we can derive the autocorrelation of $\Delta v_{Q,n}[k]$:

$$R_{\Delta Q,n}[l] = \left[\frac{a_0^2}{a_1^2} + \frac{V_M^2}{2}\cos(\Omega_I \tau)\right] R_J(\tau) \quad \text{(A.16)}$$
$$\left. * h_J(\tau) * h_J(-\tau)\right|_{\tau=lT_S}$$

where $h_J(t)$ is the impulse response of system to the aperture jitter noise

$$h_J(t) = a_1 R_D R_L T_S p_n(-t) h_E'(t). \quad \text{(A.17)}$$

Similarly, we can obtain the noise power in samples induced by the aperture jitter of EADC:

$$\sigma_{Q,n}^2 = \frac{1}{2\pi} \int_{-\infty}^{\infty} |H_J(\Omega)|^2 \left[ \frac{V_M^2}{4} S_J(\Omega + \Omega_I) \right. \\ \left. + \frac{a_0^2}{a_1^2} S_J(\Omega) + \frac{V_M^2}{4} S_J(\Omega - \Omega_I) \right] d\Omega \quad \text{(A.18)}$$

and derive its approximation as

$$\sigma_{Q,n}^2 \approx \left[ \frac{a_0^2}{a_1^2} |H_J(0)|^2 + \frac{V_M^2}{2} |H_J(\Omega_I)|^2 \right] \sigma_J^2. \quad \text{(A.19)}$$

Hence (58) is obtained.

**Feiran Su** was born in Henan, China, in 1988. He received the B.S. degree from Shanghai Jiao Tong University, China, in 2010. He is currently pursuing his Ph. D in the State Key Laboratory of Advanced Optical Communication Systems and Networks, Department of Electronic Engineering, Shanghai Jiao Tong University. His research interests include microwave photonics and high-speed analog-to-digital conversion.

**Guiling Wu** received the B.S. degree from Haer Bing Institute of Technology, Heilongjiang, China, in 1995, and the M.S. and Ph.D. degrees from Huazhong University of Science and Technology, Wuhan, China, in 1998 and 2001, respectively. He is currently a Professor with State Key Laboratory of Advanced Optical Communication Systems and Networks, Department of Electronic Engineering, Shanghai Jiao Tong University, Shanghai, China. His main research interests include photonic signal processing and transmission, and fast optical switching.

**Weiwen Zou** (S'05-M'08-SM'13) was born in Jiangxi, China, on January 3, 1981. He received the B.S. degree in physics and M.S. degree in optics from Shanghai Jiao Tong University, China, in 2002 and 2005, respectively, and Ph.D. degree in optoelectronics from the University of Tokyo, Japan, in 2008.
  In 2003, he was engaged in research of non-volatile photorefractive-based holography at University of Electro-Communications, Japan, as an exchange student. Since 2005, he had been working on Brillouin-scattering-based discriminative sensing of strain and temperature for his doctoral research in electronic engineering, the University of Tokyo, Japan. From 2008 to 2009, he was a Postdoctoral Research Fellow at the University of Tokyo. In 2009, he became a Project Assistant Professor of the University of Tokyo. In 2010, he joined Shanghai Jiao Tong University as an Associate Professor. His current research interests include fiber-optic distributed sensors, fiber-optic measurement and optical information processing.
  Dr. Zou is a member of the Optical Society of America (OSA).

**Jianping Chen** received the B.S. degree from Zhejiang University, Hangzhou, China, in 1983, and the M.S. and Ph.D. degrees from Shanghai Jiao Tong University, Shanghai, China, in 1986 and 1992, respectively. He is currently a Professor with the State Key Laboratory of Advanced Optical Communication Systems and Networks, Department of Electronic Engineering, Shanghai Jiao Tong University. His main research topics cover photonic devices and signal processing, optical networking, and sensing optics. He is also a principal scientist of 973 project in china.